\documentclass[aps,prl,preprint,nobibnotes,superscriptaddress,longbibliography]{revtex4-1}

\usepackage[breaklinks,colorlinks=true,allcolors=blue]{hyperref}
\usepackage{graphicx}
\usepackage{upgreek}
\usepackage{amssymb,amsmath,mathtools}    % need for subequations
\usepackage{mathrsfs}
\usepackage[normalem]{ulem} % for strikeout
\usepackage{xcolor} % for color

\begin{document}

\title{Internal friction controls active ciliary oscillations near the instability threshold}
	
	\author{Debasmita Mondal} 
	\affiliation{Department of Physics, Indian Institute of Science, Bangalore, Karnataka 560012, India}
	\author{Ronojoy Adhikari}
	\affiliation{The Institute of Mathematical
		Sciences\textendash Homi Bhabha National Institute, Chennai 600113,India}
	\affiliation{Department of Applied Mathematics and Theoretical Physics, Centre for Mathematical Sciences, University of Cambridge, Cambridge CB3	0WA, United Kingdom}
	\author{Prerna Sharma}
	\email{prerna@iisc.ac.in}
	\affiliation{Department of Physics, Indian Institute of Science, Bangalore, Karnataka 560012, India}

\begin{abstract}

%% 150 words

Ciliary oscillations driven by molecular motors cause fluid motion at micron scale. Stable oscillations require a substantial source of dissipation to balance the energy input of motors. Conventionally, it stems from external
fluid. We show, in contrast, that external fluid friction is negligible compared to internal elastic stress through a simultaneous measurement of motion and flow field of an isolated and active \textit{Chlamydomonas} cilium beating
near the instability threshold. Consequently, internal friction emerges as the sole source of dissipation for ciliary oscillations. We combine these experimental insights with theoretical modeling of active filaments to show that an instability to oscillations takes place when active stresses are strain softening and shear thinning. Together, our results reveal a counterintuitive mechanism of ciliary beating and provide a general experimental and theoretical methodology to analyze other active filaments, both biological and synthetic ones.

\end{abstract}
	
\maketitle

%%%% for making Figure & Table caption label in bold
\makeatletter
\renewcommand*{\fnum@figure}{\textbf{Fig.~\thefigure}}
\renewcommand*{\fnum@table}{\textbf{Table~\thetable}}
\makeatother

\section*{Introduction}

Cilia and flagella are prototypical engines of micron-scale motility, used by the biological world in myriad contexts  \cite{Goldstein2015Rev,Gaffney2011mammalian}. They are classic examples of nonequilibrium active filaments that undergo
spontaneous oscillations by converting stored or ambient energy into mechanical motion \cite{Sriam2013RevModPhys}. Their elemental structure, called axoneme, consists of cross-linked microtubule (MT) doublets and dynein motors which apply forces on MT, through adenosine triphosphate (ATP) hydrolysis, to cause periodic bending of the whole structure \cite{Nicastro2006}.
In addition to its importance in the cellular context, ciliary beating has been mimicked in synthetic filaments for applications in cargo transport, microfluidics and drug delivery \cite{Dreyfus2005microscopic,Paxton2004catalytic}. 

Naturally, there have been a number of studies devoted to understand the microscopic mechanisms of ciliary oscillations \cite{Brokaw1971SlidingFilament,Rikmenspoel1978Sperm,HinesBlum1979Rev,Riedel-Kruse2007,Gaffney2011mammalian,BaylyWilson2015Stability,Sartori2016DynamicCurv}. Most of them analyze beating based on passive elastic stresses calculated from measured filament waveform, models of active drive and dissipation stemming from external fluid. The dissipation is usually computed using slender body integral equations since filament waveform alone cannot be used to validate its form. Intuitively, hydrodynamics should play an essential role as cilia operate in the regime of low Reynolds number ($ \sim 10^{-3} $). However, its contribution to synchronization and collective behavior of cilia is highly debated because most of these in vivo experiments are conducted on live cells where it is difficult to delineate hydrodynamics from other coupling mechanisms such as cell body rocking or intracellular means \cite{GoldsteinSynchro2009,GeyerSynchroRocking2013,TamSynchro2015,Wan2016coordinated,KlindtFriedrichPRL2016LoadResponseCR,Elgeti2013emergence}. In addition, biopolymers often have substantial internal friction \cite{MarkoPoirier}. Hence, to determine the dominance or lack of hydrodynamics in ciliary oscillations, an accurate measurement of the external viscous drag and its competition with the other elastic stresses for an in vitro system of cell-free isolated cilium is required, where cell body rocking and intracellular basal body coupling are eliminated. This can be accomplished by measuring its flow field along with the waveform.

Here, we present the first simultaneous measurement of
the bending waveform and flow field of an isolated and active \textit{Chlamydomonas} axoneme, beating near the critical ATP concentration at which oscillations set in, at high spatio-temporal resolution to address the role of external fluid in its beating. Our measurements demonstrate that hydrodynamic dissipation, accurately described by resistive force theory (RFT), is negligible compared to internal elastic stresses. Consequently, a dissipation mechanism internal to the filament is essential for stable driven oscillations, in contrast to the widely-held view that fluid friction is the only important source of dissipation \cite{HinesBlum1979Rev,Sartori2016DynamicCurv}. We combine these insights with a theoretical model of filament motion that includes a minimal spring-dashpot form of active stress. We show there exist critical values of active stress beyond which the model exhibits oscillations, namely, active stresses should be strain softening and shear thinning.

\section*{Results}

\subsection*{{\normalsize Experimental system}}

\begin{figure}[t]
	\centering 
	\includegraphics[width=0.8\linewidth]{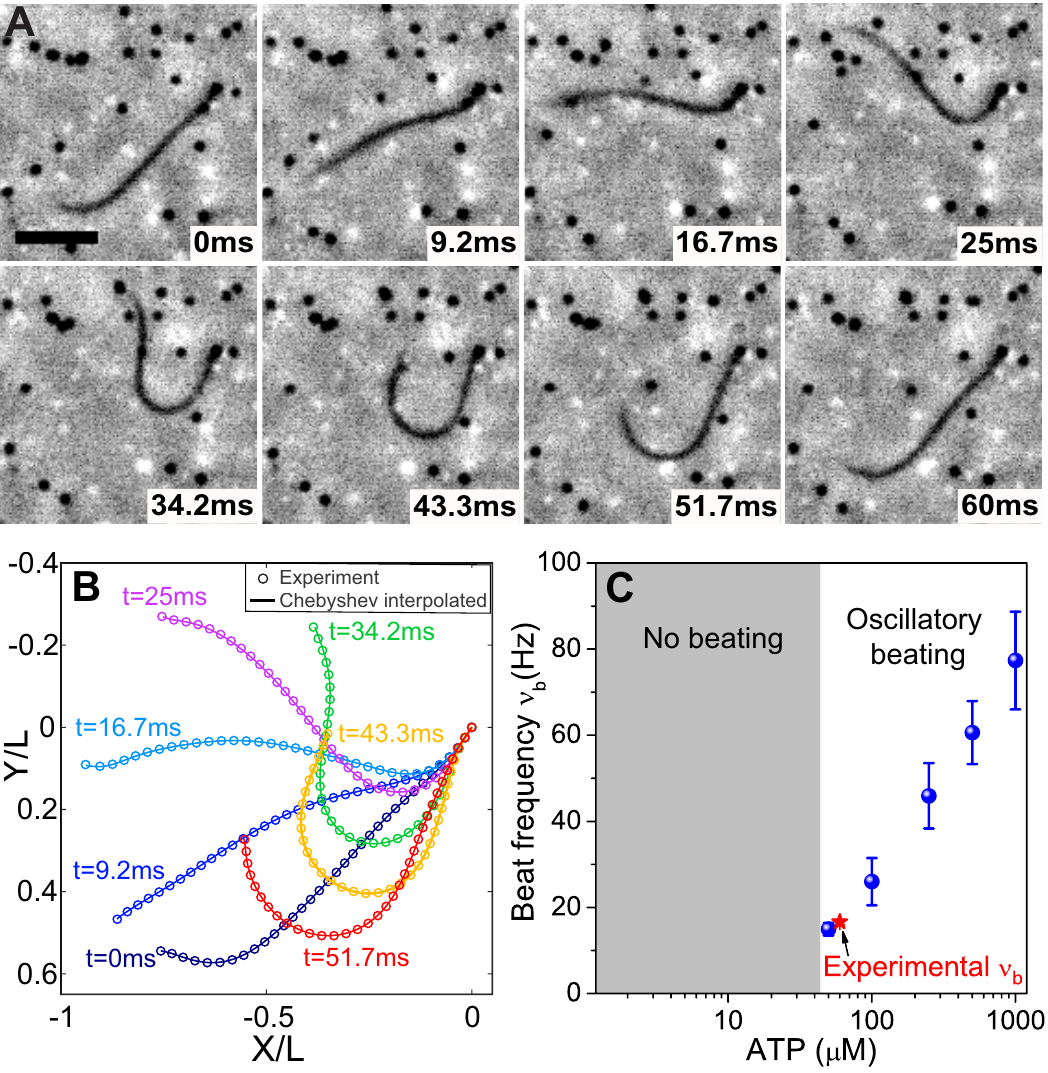} 
	\caption{\textbf{Simultaneous measurement of motion and flow.} (\textbf{A}) Time lapse snapshots of a clamped and reactivated \textit{Chlamydomonas} axoneme at 60 $\upmu$M ATP in the presence of tracer particles (black dots). Scale bar, 4.2 $\upmu$m. (\textbf{B}) Nondimensionalized experimental positions  and their corresponding Chebyshev interpolants. (\textbf{C}) Beat frequency of axonemes as a function of ATP concentration  (blue circles). Red star indicates experimental conditions that are near the threshold for the onset of oscillations.}
	\label{Figure1-Motion} 
\end{figure}

Isolated and reactivated axonemes of $ \sim 11 ~\upmu$m length and $ \sim 0.2 ~\upmu$m diameter \cite{Nicastro2006}, purified from
the unicellular algae \textit{Chlamydomonas reinhardtii}, are clamped at one end on a glass cover slip. Their oscillations, with beat frequency $ \nu_b \sim 16.22 $ Hz, are approximately planar with an average centreline height $h \sim 0.9 ~\upmu$m from the surface. Passive microspheres
are introduced into the suspension as tracers for measuring flow of the ambient fluid using particle tracking
velocimetry (PTV) (see Materials and Methods). We capture motion of both the axoneme and tracers at high spatio-temporal resolution using a 60X phase objective coupled with a high speed camera (Fig.~\ref{Figure1-Motion}A and movie S1). The detailed experimental procedure is described in Material and Methods. The position of the axoneme is sampled at many points
over its length to construct a global Chebyshev polynomial interpolant of the parametric form $\boldsymbol{R}(s)=\sum\boldsymbol{a}_{n}(t)T_{n}(s)$
where $0\leq s\leq L$ is the arc-length and $\boldsymbol{a}_{n}$
is a vector of Chebyshev coefficients for the $x$ and $y$ components
of the position (Fig.~\ref{Figure1-Motion}B).

\subsection*{{\normalsize Filament geometry and mechanics}}

We perform experiments near the critical ATP concentration at which the axoneme transitions from quiescent to oscillatory state through an instability threshold (Fig.~\ref{Figure1-Motion}C) \cite{CamaletJulicher2000}. Near this threshold, details of the internal structure of the axoneme become irrelevant.  Therefore, we model it as an inextensible, but shearable, active rod of uniform circular cross section of diameter $a$ and length $L$, with a centreline described by the curve $\boldsymbol{R}(s)$ to calculate stresses averaged over its cross-section. We attach an orthogonal Frenet-Serret frame to each point of the curve. Planar motion of the filament results in the curve tangent to be parametrised by the tangent angle $ \theta $ as $\boldsymbol{t}(s)=[\cos\theta(s),\sin\theta(s)]$, which automatically satisfies the inextensibility constraint, $\boldsymbol{t}\cdot\boldsymbol{t}=1$. The position of the curve is obtained in terms of the tangent angle by integrating
$\partial_{s}\boldsymbol{R}=\boldsymbol{t}$. The shear strain of
the rod is $u(s)$ which is assumed to vanish at the clamped base, $u(0)=0$. Inextensibility then requires the shear strain to accommodate the filament bending as $u(s)=a[\theta(s)-\theta(0)]\equiv a\Delta\theta(s)$
and the kinematics of the filament is completely specified by  $\theta$ or equivalently by shear angle, $\Delta \theta$ \cite{Brokaw1971SlidingFilament}. Notably, the unobservable shearing of the filament
can be inferred from its observable curvature, $\kappa=\partial_{s}\theta \equiv \partial_{s}\Delta\theta$. There is no assumption of small curvature in this kinematic description. Internal active stresses cause the filament to shear and, by the kinematic constraint, to curve.

We assume that the rod supports an internal stress whose stress and
moment resultants on the cross section
at $s$ are, respectively, $\boldsymbol{F}(s)$ and $\boldsymbol{M}(s)$
and that it is acted upon by forces and moments whose sum per unit
length are, respectively, $\boldsymbol{f}$ and $\boldsymbol{m}$.
Then, in the absence of inertia, the balance equations for force and
torque are \cite{ericksen} 
\begin{equation}\label{basic_FBMB_eqn}
\partial_{s}\boldsymbol{F}+\boldsymbol{f}=0,\quad\quad\partial_{s}\boldsymbol{M}+\boldsymbol{t}\times\boldsymbol{F}+\boldsymbol{m}=0
\end{equation}
Internal moments included in $\boldsymbol{m}$, which can exist only if the internal stress is antisymmetric, are generally omitted in the elasticity of rods, but are essential to our model \cite{ericksen}. The above equations are closed by
identifying the relevant forces, moments and their
constitutive equations in terms of the kinematic variables. Integrating
the force equation, the stress resultant can be expressed as $\boldsymbol{F}(s)=\int_{s}^{L}\boldsymbol{f}(s) \thinspace ds+\boldsymbol{F}(L)$, where
$\boldsymbol{F}(L)=0$ at the
free terminus of the rod. The only force per unit length relevant
here is the external hydrodynamic drag, $\boldsymbol{f}^{v}$.

\subsection*{{\normalsize External viscous drag}}

A rod moving through a viscous fluid experiences a drag $\boldsymbol{f}^{v}(s)$
and creates a flow $\boldsymbol{v}(\boldsymbol{r})$.
In the experimentally
relevant limit of slow viscous flow and a slender rod, $L\gg a,$
the integral representation of Stokes equation gives 
\begin{equation} \label{Eq:exterior_flow}
\boldsymbol{v}(\boldsymbol{r})=-\int_{0}^{L}\boldsymbol{G}(\boldsymbol{r},\boldsymbol{R}(s))\cdot\boldsymbol{f}^{v}(s)\thinspace ds
\end{equation}
This represents a distribution of Stokeslets of strength $\boldsymbol{f}^{v}(s)$ where $\boldsymbol{G}$ is a Green's function of the Stokes equation.
The matching of the fluid flow to the velocity of the rod at its surface
yields the slender body integral equation whose formal solution is $\boldsymbol{f}^{v}(s)=-\int_{0}^{L}\boldsymbol{\gamma}(s,s')\cdot\dot{\boldsymbol{R}}(s')\thinspace ds'$. 
The friction kernel is often approximated by a local form
with a constant friction coefficient, $\boldsymbol{\gamma}(s,s')=\boldsymbol{\gamma}\delta(s-s')$.
In this RFT limit, the drag is 
\begin{equation} \label{hydro_RFT-1}
\boldsymbol{f}^{v}(s)=-\boldsymbol{\gamma}\cdot\dot{\boldsymbol{R}}(s),\quad\boldsymbol{\gamma}=\gamma_{n}\boldsymbol{n}\otimes \boldsymbol{n} + \gamma_{t}\boldsymbol{t} \otimes\boldsymbol{t}
\end{equation}
where $\gamma_{n}$ and $\gamma_{t}$ are the normal and tangential components of the friction coefficient and $ \dot{\boldsymbol{R}}=\dot{R}_n \boldsymbol{n} + \dot{R}_t\boldsymbol{t}$ is the centreline velocity of the rod in terms of its normal and tangential components.

\begin{figure}[h]
	\centering 
	\includegraphics[width=0.9\linewidth]{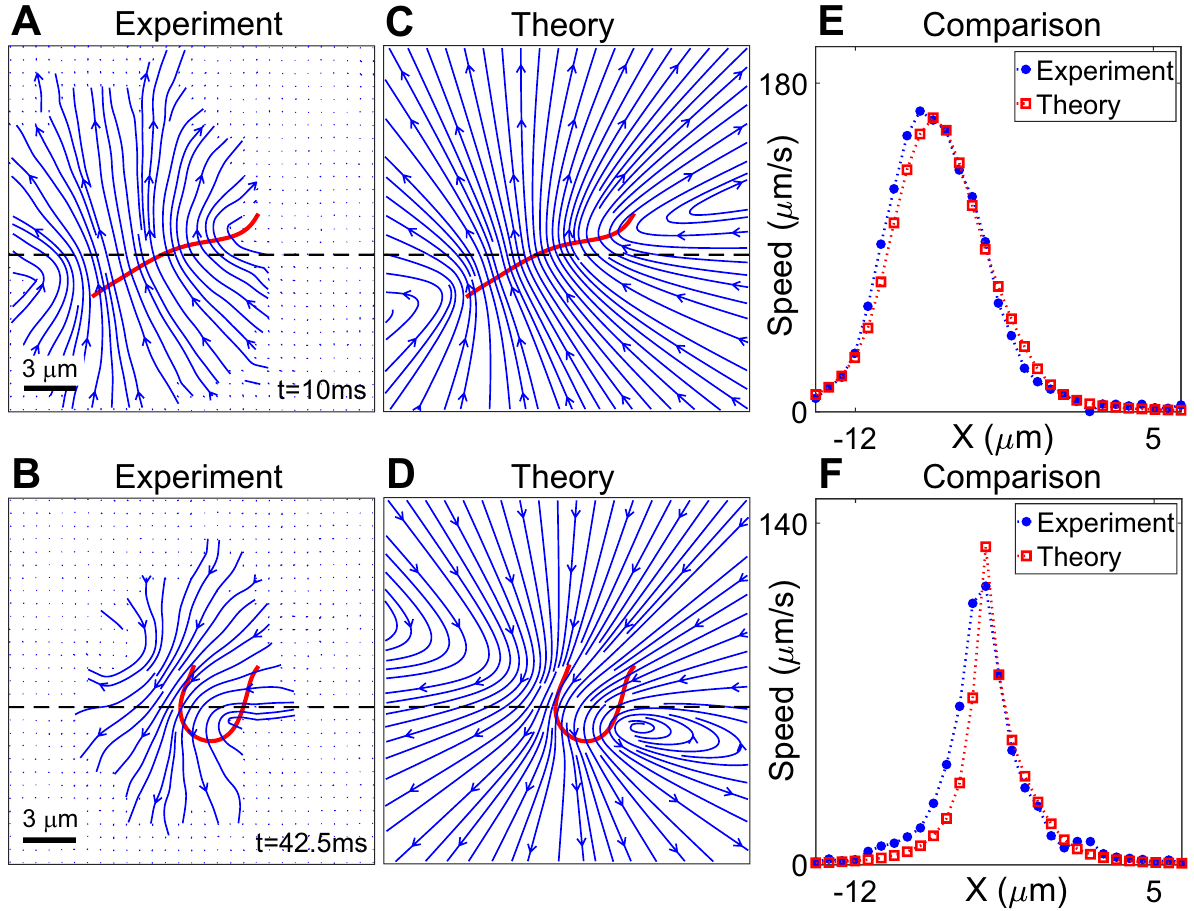}
	\caption{\textbf{External drag force.} Experimentally measured instantaneous flow fields using PTV  during (\textbf{A}) power and (\textbf{B}) recovery stroke for the corresponding initial axonemal conformations (red lines). Streamlines represent flows with  $v > 8~\upmu$m/s and vector fields for lower speeds in the Brownian noise regime. (\textbf{C} and \textbf{D}) Theoretically computed flow fields using RFT corresponding to (A) and (B) respectively. (\textbf{E} and \textbf{F}) Comparison of experimental and theoretical flow fields along representative cuts (dashed lines).}
	\label{Figure2-Flowfield} 
\end{figure}

With the viscous drag thus determined in terms of the centreline velocity, we now validate RFT using experimentally measured instantaneous flow fields (Fig.~\ref{Figure2-Flowfield}, A and B). We compare these experimental flows with theoretically computed ones using  Eqs. \ref{Eq:exterior_flow} and  \ref{hydro_RFT-1} (Fig.~\ref{Figure2-Flowfield}, C and D), wherein  $ \dot{\boldsymbol{R}}$ is determined from the measured waveform, $ \gamma_n=4\pi\eta/ln(4h/a)=4.35$ mPa.s, $\gamma_t=\gamma_n/2 $  \cite{Howard2001mechanics} (fluid viscosity, $ \eta=1 $ mPa.s) and $\boldsymbol{G}$ is the Lorentz-Blake tensor for flow near a no-slip wall \cite{blake1971}. Representative cuts along the experimental and theoretical flows show that there is a good agreement between the two (Fig.~\ref{Figure2-Flowfield}, E and F).
A more comprehensive comparison is  given by root mean square deviation of the flows, $RMSD = \sqrt{\sum_{i=1}^{NS} (v^{expt}_i - v^{th}_i)^2/NS}$, where $NS$ is the number of grid points, $v^{expt}_i$ and $v^{th}_i$ are the experimental and theoretical flow magnitudes at the $i$-th grid point, respectively. RMSD of $v_x$, $v_y$ and $|\boldsymbol{v}|$ in Fig.~\ref{Figure2-Flowfield} are (A and C) 4.32, 8.28 and 7.47; (B and D) 7.26, 7.89 and 9.81 $ \upmu $m/s,  respectively,  all of which are within the Brownian noise regime implying a good match. Therefore, we have now verified by direct measurement that the hydrodynamic drag force is unambiguously given by RFT. This form of drag is used to evaluate the stress resultant, $\boldsymbol{F}$ and its normal component, $ F_n $, is used in the torque balance equation below.

\subsection*{{\normalsize Scalar equation of motion of the filament}}

\begin{figure}[h]
	\centering 
	\includegraphics[width=\linewidth]{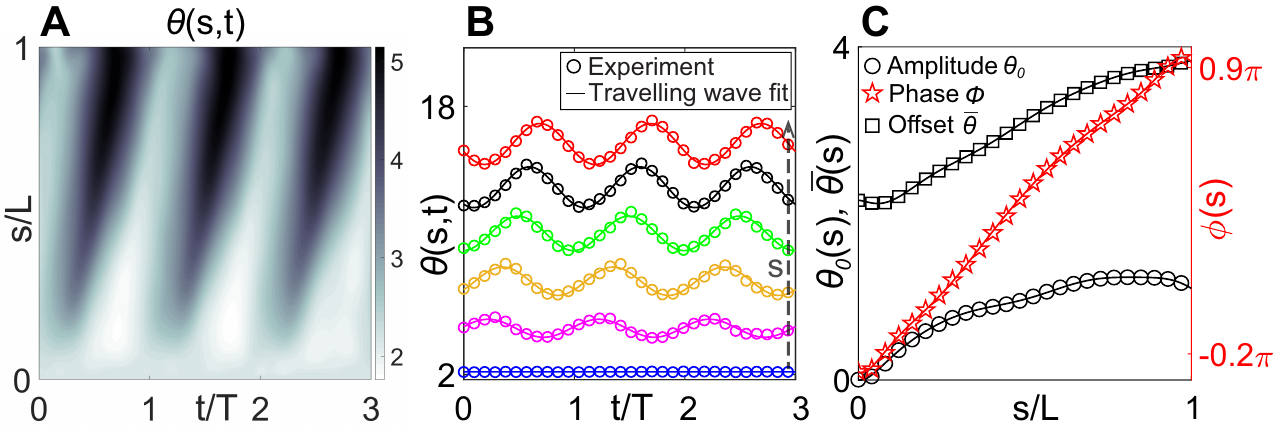} 
	\caption{\textbf{Traveling wave parametrization to tangent angle.} (\textbf{A}) Space-time plot of the experimental $\theta$. (\textbf{B}) Traveling wave fit to experimental $\theta$ at representative  arc-lengths, $ s $. Plot for $\theta$ at each s is shifted along the $ y $ axis, for clarity. (\textbf{C}) Amplitude, phase and offset of the traveling wave in $\theta$ along $ s $ and their corresponding Chebyshev interpolants (solid lines).}
	\label{Figure3-TravellingWave} 
\end{figure}

In slow viscous flow as is shown above, the drag acts in the plane of motion and the stress resultant remains in that plane. Therefore, the couple resultant $ \boldsymbol{M}$ is normal to the plane of motion, along the frame binormal $ \boldsymbol{b} $, and vanishes at the free terminus, giving $M(L)=0$. As the only non-zero components of $\boldsymbol{t}\times\boldsymbol{F}=F_n\boldsymbol{b}$ and $\boldsymbol{m}$ are normal to the plane, torque balance reduces to a scalar equation, $\partial_s M + F_n +m=0$. The dissipation in this equation is contained in the normal component of the stress resultant due to external viscous drag as $ F_n(s) = \boldsymbol{n}(s) \cdot \int_s^L \boldsymbol{f}^v(s') ds'=-\gamma_ng'_n(s)$ where $ g'_n(s) = \boldsymbol{n}(s) \cdot \int_s^L \big[\dot{R}_n(s')\boldsymbol{n}(s') + \dot{R}_t(s')\boldsymbol{t}(s')/2 \big] ds' $ using Eq. \ref{hydro_RFT-1}. Herein, the filament velocity components can be expressed in terms of the tangent angle, $ \theta $,  using $ \dot{\boldsymbol{R}}(s) = \int_0^s \dot{\boldsymbol{t}}(s')ds' $. The simplest elastic contributions for an inextensible and shearable active rod are that due to bending, $ M=EI\kappa $, and shear, $ m=-aku $  \cite{Brokaw1971SlidingFilament}. The source of filament motion is included as the  active  moment per unit length, $ m^\mathcal{A}$. Material parameters in these constitutive relations 
are bending rigidity, $EI=  600 $ pN$\upmu$m$^2 $ \cite{XuDutcher2016flexural,Sartori2016DynamicCurv} and shear stiffness, $k= 2000$ pN/$\upmu$m$^2$ \cite{Minoura2001Spring,XuDutcher2016flexural}.  Torque balance, closed by the constitutive
relations and the expression for $ F_n $, yields the
following dynamical equation,
\begin{equation}\label{torque_eqn1}
EI\partial_s \kappa -aku -\gamma_ng'_n  + m^\mathcal{A}=0
\end{equation}
All the passive terms in this equation can be expressed completely in terms of the tangent angle, $ \theta $. Experimental data reveals that the tangent angle can be  parametrised by a traveling waveform,
$\theta(s,t)=\theta_{0}(s)\sin[\omega t-\phi(s)]+\bar{\theta}(s)$,
where $\theta_{0}$, $\omega$, $\phi$ and $\bar{\theta}$ are the amplitude,
angular frequency, phase and offset respectively (Fig.~\ref{Figure3-TravellingWave}) \cite{Rikmenspoel1978Sperm,Sartori2016DynamicCurv}. As we are interested in oscillations about the time averaged shape of the beat, $\bar{\theta}$, we define $ \theta'(s,t)=\theta(s,t)-\bar{\theta}(s) $. In the following, we focus on $ \theta' $ and drop the prime such that $ \theta(s,t) \equiv \theta_{0}(s)\sin[\omega t-\phi(s)]  $ which represents the dynamic oscillatory beat of the filament about the mean shape. Therefore, we use this parametric form, instead of the Chebyshev interpolant, to estimate all angle-dependent quantities in the dynamical equation.

The dimensionless equation of motion with $ l_\kappa = \sqrt{EI/a^2k}=2.74 ~\upmu $m as the curvature penetration length scale and $ \nu_h = EI/\gamma_n l_\kappa^4 = 2447 $ Hz as the hydrodynamic relaxation frequency scale of the system is,
\begin{equation}\label{torque_eqn1_ND}
\partial_s^2\Delta\theta - \Delta\theta - g_n + m^\mathcal{A} = 0
\end{equation}
Here, $ g_n = g'_n/l_\kappa^2 \nu_h $ and $ m^\mathcal{A} $ is rescaled by $ l_\kappa^2 /EI $. Using $ \theta $ from the traveling wave parametrisation of experimental data and constitutive parameters from the literature, the space-time variation of elastic and viscous terms of Eq. \ref{torque_eqn1_ND} are plotted in Fig.~\ref{Figure4-DynamicsStability} (A to C). It shows that the nonlinear viscous dissipation has a standing wave component in contrast to the linear elastic terms \cite{Rikmenspoel1978Sperm}.
On comparing their colorbars, we conclude that the hydrodynamic dissipation, $ g_n \sim \mathcal{O}$(0.01), is negligible compared to the elastic forces which are of $ \mathcal{O} $(1).

The axoneme consumes energy in the form of ATP and exhibits stable oscillations. Such a continuously driven system can be in a dynamical steady state only when the elastic stresses due to the drive are balanced by some dissipation. Since external fluid friction cannot account for this dissipation, consistency demands that the internal stress has, in addition to an elastic component, a dissipative frictional component. Each kinematic degree of freedom, i.e. bending and sliding, can contribute to dissipation. Notably, bending friction in MTs has been experimentally demonstrated to be dominant over hydrodyamic friction for length scales smaller than  $20 ~\upmu$m and attributed to slow structural rearrangements \cite{MarkoPoirier,BrangwyneDogicInternalViscosity,Janson2004}. The bending friction coefficient of a 11 $\upmu$m long axoneme, $\Gamma_\kappa = 1.6~\rm{pN} \upmu \rm{m^2 s} $, is same as that of a single MT since it is an intensive quantity \cite{BrangwyneDogicInternalViscosity,Janson2004}. Though there is no experimental measure of the shear friction coefficient of an axoneme, $ \Gamma_u$, several experimental studies suggest the presence of inter-MT sliding friction \cite{Minoura2001Spring}. We consider $ \Gamma_u = 10~ \rm{pN s/}\upmu \rm{m} $ from estimates of nexin protein friction \cite{riedel2005mechanics} (see table \ref{Table:SlidingFriction}). Earlier work introduced similar terms for internal viscous stresses for either stabilization of the numerical simulations of  bend propagation in active filaments \cite{Brokaw1972computer,murase1986model,HinesBlum1978bend,HinesBlum1979Rev} or on the basis of theory alone \cite{BaylyWilson2015Stability,BaylyDutcher2016Flutter}. However, here we have shown experimentally that such terms are necessary to completely account for dissipation in ciliary motion.

\subsection*{{\normalsize Dynamical equation without external friction}}

We, therefore, neglect the external viscous drag and include  the internal viscous stresses to re-write the scalar torque balance equation, $ \partial_s M +m=0 $. Hence, the constitutive equations are modified as  $ M=EI\kappa +\Gamma_\kappa \dot{\kappa} $ and $ m=-aku-\Gamma_u\dot{u} +m^\mathcal{A} $. The modified dynamical equation in terms of the shear angle, $ \Delta\theta $, is $EI\partial_s^2\Delta\theta +\Gamma_\kappa \partial_s^2 \partial_t\Delta\theta -a^2 k \Delta\theta -a\Gamma_u \partial_t\Delta\theta +m^\mathcal{A} =0$. The negligible role of fluid friction leads to a second-order reaction-diffusion equation for the shear angle, rather than  fourth-order partial differential equations that are commonly obtained when fluid friction is retained \cite{BaylyWilson2015Stability,Riedel-Kruse2007,murase1986model}.
Identifying the two frequency scales of the system, a curvature relaxation frequency scale, $ \nu_\kappa = EI/\Gamma_\kappa $ and a sliding relaxation frequency scale, $ \nu_u= ak/\Gamma_u $, we note that $ \nu_\kappa/\nu_u =9.38 $ i.e both kinematical degrees of freedom contribute to dissipation. The dimensionless equation of motion with $ \nu_\kappa$ as the frequency scale is,
\begin{equation} \label{torque_eqn2_ND} 
\partial_s^2\Delta\theta + \partial_s^2 \partial_t\Delta\theta - \Delta\theta - \frac{\nu_\kappa}{\nu_u}\partial_t\Delta\theta +m^\mathcal{A} =0
\end{equation}
where $ m^\mathcal{A} $ is rescaled by $ l_\kappa^2 /EI $.
Figure~\ref{Figure4-DynamicsStability} (D and E) show the variation of the internal viscous stresses over three beat cycles along the filament length. The colorbars in
Fig.~\ref{Figure4-DynamicsStability} (A to E) indicate that the internal viscous stresses due to bending friction of $ \mathcal{O}(0.1) $ and shear friction of $\mathcal{O}(1)$ compete with the elastic stresses of $\mathcal{O}(1)$, unlike the negligibly small external drag of $ \mathcal{O}(0.01)$.

The internal passive stresses being completely defined, a constitutive relation for the active moment in terms of kinematic variables is required. The axoneme having both elastic and viscous material parameters, the motor activity will induce a viscoelastic response. Hence, we assume a minimal spring-dashpot form for the dynamics of $ m^\mathcal{A} $, $\partial_t m^\mathcal{A}+b_1m^\mathcal{A}=b_3\Delta\theta $ \cite{CamaletJulicherProst1999,CamaletJulicher2000}, whose Fourier representation describes the spring-like elastic and dashpot-like viscous dampening response of the system under oscillating shear as shown later. Here, the active stress is parametrised by two constants: $ b_1 $ controls the autonomous dynamics of $ m^\mathcal{A} $ and $ b_3 $ controls the amount of feedback it receives from the sliding kinematics. We have chosen this simplest form of constitutive relation as it is first order in time, lowest order in wavenumber and linear in $ \Delta\theta $. This minimal constitutive relation is most relevant for a coarse-grained description of axoneme as used here. This relation for $ m^\mathcal{A} $ together with Eq. \ref{torque_eqn2_ND} form a pair of coupled equations, 
\begin{equation}\label{coupled_eqn}
\partial_t 
\begin{bmatrix}
(\partial_s^2 - \nu_\kappa/\nu_u) \Delta\theta \\ m^\mathcal{A}
\end{bmatrix}
=
\begin{bmatrix}
1-\partial_s^2  & -1  \\
b_3 & -b_1 
\end{bmatrix}
\begin{bmatrix}
\Delta\theta\\ m^\mathcal{A}
\end{bmatrix}
\end{equation}
which emphasizes that the passive and active parts in this model are independent, but coupled, degrees of freedom. 

These dissipative linear partial differential equations can sustain stable oscillations only in the presence of non-linearities whose choice will become crucial far from the instability threshold \cite{CamaletJulicher2000,Oriola2017Nonlinear,BaylyWilson2015Stability}. We now focus on the linear regime and seek to identify the threshold for the onset of oscillations and its frequency near threshold. This is relevant to our experimental results because the axoneme in this study beats at 60 $\upmu$M ATP, near the critical ATP concentration of 45 $\upmu$M at which oscillations set in (Fig.~\ref{Figure1-Motion}C). Hence, the axoneme is beating near the instability threshold, where the nonlinearity is weak and the oscillation frequency of the limit cycle is that of the linear analysis evaluated at the threshold \cite{CamaletJulicher2000,Oriola2017Nonlinear}.

We perform linear stability analysis on the coupled Eq. \ref{coupled_eqn} in the Fourier domain (see section S2 for detailed analysis). The dispersion relation is quadratic in complex frequency $ z $, $\big[(1+q_n^2)-iz (q_n^2+ \nu_\kappa/\nu_u)\big](iz-b_1) + b_3 = 0$. The clamped head-free end boundary conditions of the filament discretize the wavenumbers to $ q_n = \dfrac{n\pi}{2(L/l_\kappa)} $ for odd $ n $. Unstable oscillations are possible only when both $b_1<0$ and $b_3<0 $ (Fig.~\ref{Figure4-DynamicsStability}F).
These parameters can alternatively be related to the elastic, $ G' $, and viscous, $ G'' $, response coefficients of the active stress through its fundamental Fourier mode of oscillation frequency $ \omega $ as $ \widetilde{m^\mathcal{A}}= (G'+i\omega G'')\widetilde{\Delta\theta}$, where $ G'= b_1b_3/(b_1^2+\omega^2)$ and $G''=-b_3/(b_1^2+\omega^2) $ \cite{Riedel-Kruse2007,CamaletJulicher2000}.
The sign of response coefficients determines the nature of active stress. If $G', G''<0 $, the system's passive spring constant and friction coefficient get renormalized by the ATP dependent dynein activity. In our case, $ b_1,b_3<0 \implies G', G''>0$, i.e. activity works against the material response and leads to strain softening ($ G'>0 $) and shear thinning ($ G''>0 $) in the system (see section S3 for detailed discussion).
The experimental beat frequency of the axoneme ($ \omega_{expt}=2\pi \nu_b/\nu_\kappa = 0.272 $), operating close to the threshold, constrains the parameters of the constitutive relation at hand, namely, $ b_1=-0.121,~b_3=-0.85 $ for the fundamental oscillatory mode (Fig.~\ref{Figure4-DynamicsStability}G). Elastoviscous response coefficients of the active stress computed with these values, $ G' = 1.16,~G''=9.6 $, imply that the viscous response of the active stress dominates the elastic one in our experiment. On comparing the nature of active stress so obtained with a microscopic model of load dependent detachment of motors in the experimentally relevant dominantly linear regime of the post-threshold dynamics \cite{Riedel-Kruse2007,CamaletJulicher2000,Oriola2017Nonlinear}, we infer that axonemal dyneins are low duty ratio  motors (see section S4 for details), which agrees with previous studies \cite{Howard2001mechanics}.

We emphasize again that the existence and dominance of internal friction over hydrodynamic drag in isolated ciliary dynamics is borne out of experimental measurements alone and not through detailed modeling. Simultaneous measurement of flow field and waveform of an active cilium gives us the unambiguous measure of the hydrodynamic drag force i.e. the external fluid friction. The passive internal elastic stresses are calculated from the experimental waveform using minimal constitutive relations for bending and shear elasticity which are widely accepted in the literature for elasticity of rods. Comparing the experimentally measured fluid friction with that of the passive elasticity led to the conclusion that fluid friction is not enough to counteract the elastic stresses due to the active drive and consequently, stable ciliary oscillations need internal friction to reach their dynamical steady state. The above theoretical analysis of the filament equation of motion using a minimal constitutive relation for the active drive is simply to illustrate that oscillations exist in presence of internal friction too.

\section*{Discussion}

In conclusion, to the best of our knowledge, this is the first direct experimental evidence of the negligible role of external fluid friction in ciliary oscillations near the instability threshold, thereby suggesting that internal dissipation mechanisms are essential to have a self-consistent understanding of ciliary beating. Our results are complementary to studies conducted on live cells/sperms \cite{QinGopinath2015SciRep} and reactivated cilia at high ATP conditions \cite{Brokaw1975Viscosity} wherein changing the fluid viscosity alters both waveform and beat frequency of cilia by about 20\%. At low ATP concentration investigated in this study, ciliary beat frequency and waveform is almost independent of the viscosity of ambient fluid \cite{Brokaw1975Viscosity}. It is crucial to study this regime near the instability threshold because, here, the system is governed by linear terms that are generic and do not depend on the minor structural details of the system, leading to an understanding that is universal in nature \cite{CamaletJulicher2000}. As in this study, our results imply that one should include the contribution of internal friction with the obvious source of external fluid friction in constructing balance equations for active filaments in viscous fluids to correctly understand the dynamical regimes in which they operate.

This result will inspire further studies on the role of fluid friction in ciliary beating far from the threshold and in collective behavior of cilia. In addition, the measured flow field can distinguish internally active filaments from those driven by surface forces, for example, in phoretic chains (see fig.~S1) \cite{LaskarAdhikari2013SciRep}. Therefore, a measurement of the flow field of an oscillating cilium provides vital information on its mechanisms of operation that cannot be obtained from the measurement of motion alone. Our forward approach of using the experimental insights from simultaneous waveform and flow measurements to build a theoretical model of active filament is in contrast to the existing reverse approaches of starting with microscopic models of active stress to match the observed macroscopic waveform of the filament \cite{Sartori2016DynamicCurv,BaylyWilson2015Stability,Riedel-Kruse2007}. Lastly, our approach can serve as a paradigm for analysis and regulation of any active slender body, both biological and synthetic one, in a viscous fluid.

\section*{Materials and Methods}

\textbf{Axoneme purification and reactivation:} 
Wild-type \textit{Chlamydomonas reinhardtii} cells (strain: CC1690) are synchronously grown in 12:12 hours light:dark cycle in TAP+P medium (tris acetate phosphate) \cite{ALPER2013}. They are collected 2-3 hrs after the beginning of light cycle, at OD$_{750}$ (optical density at 750 nm) $\approx 0.15-0.25$, followed by washing in HES buffer [10 mM HEPES (pH 7.4), 1 mM EGTA, 4\% Sucrose] thrice. The plasma membrane covering the cells is disrupted by adding 0.15\% of a non-ionic detergent  IGEPAL CA-630 (I8896, Sigma-Aldrich)  in HMDEK buffer [30 mM HEPES (pH 7.4), 5 mM MgSO$_4$, 1 mM dithiothreitol (DTT), 1 mM EGTA, 50 mM K-acetate] to the cell pellet. The cells in IGEPAL are then kept in ice for $ \sim 5-7 $ hours. These cells are devoid of cell membranes and are called cell models. Some of them shed their demembranated flagella, called axonemes, due to weakening of the attachment to the cell body. Isolated axonemes are then separated from the cell models by centrifugation. Axonemes are then mixed with 30\% saturated sucrose, flash freezed and stored at -80$^{\circ}$C for long term usage. 

This method of purification, by long-term exposure to detergent, yields nonsticky axonemes, in contrast to the commonly used dibucaine procedure \cite{ALPER2013,WAKABAYASHI2015},  hence essential for flow field measurements. These non-motile axonemes regain their motility in presence of 45 $\upmu$M to 1 mM ATP in HMDEKP buffer (HMDEK + 1\% PEG-20k) and are said to be reactivated. There was no distinction in the nature of beating of the frozen axonemes, when thawed, from those that were not frozen. We use an ATP regeneration system, to hold the ATP concentration constant within the sample for approximately 40 mins, enough for imaging approximately 6-7 isolated axonemes in one sample. The ATP regeneration system composes of 6 mM sodium creatine phosphate (27920, Sigma-Aldrich) and 40 U/ml creatine phosphokinase (CPK) from bovine heart (C7886, Sigma-Aldrich).
The commercially available CPK is typically oxidized, hence we reduce it by DTT at room temperature for efficient reactivation \cite{Favre2005CPK}. 

\vskip 2.5mm

\textbf{Tracer particles:} 
The tracer particles chosen, for flow field measurements, are neutrally buoyant (polystyrene microspheres with a density of $ \rm 1.055~ g/cc$) and small (diameter of 200 nm, the lowest size that can be used with diffraction-limited optical imaging) so that their motion is nearly identical to the fluid in which they reside. Commercially available charge-stabilized microspheres stick to each other and to the axoneme due to screening of electrostatic interactions by the divalent ions and salts present in the reactivation buffer. We graft long chains of block copolymer PLL-PEG20k [poly-L-lysine (PLL), P7890, 15 to 30 kDa, Sigma-Aldrich; mPEG-SVA-20k, NANOCS], 
onto 200-nm negatively charged sulfate latex beads (S37491,Thermo Fisher Scientific) to impart additional steric stabilization. 

\vskip 2.5mm

\textbf{Surface modification of glass surfaces and sample preparation:} 
The coverslips and slides are cleaned with a hot soap solution (1\% Hellmanex III), followed by rinsing with ethanol and 100 mM potassium hydroxide. We graft polyacrylamide brush on these clean glass surfaces to suppress depletion interaction of beads and filament with the surfaces. The coverslips are further modified with randomly located sticky patches of 0.05\% PLL over the polyacrylamide brush, to clamp the axonemes at one end. We introduce the reactivation buffer containing axonemes and tracer particles inside a sample chamber made up of a glass slide and coverslip sandwich with double sided tape of thickness 65 $\upmu$m as the spacer. 

\vskip 2.5mm

\textbf{Imaging and depth of focus:}
The sample is mounted on an inverted microscope (Olympus IX83), equipped with 60x oil immersion phase objective [0.65 to 1.25 numerical aperture (NA), UPlanFL N, PH3] and connected to a high speed complementary metal oxide semiconductor (CMOS) camera
(frame rate, 1200 frames per second; Phantom Miro C110, Vision Research) for imaging. We use an intermediate NA between 0.65 and 1.25 for the 60x variable NA phase objective, to capture most of the filament beat in focus. We have measured the depth of focus ($ \Delta Z $) of the objective at this intermediate NA to be $\rm 1.4~ \upmu m $. We only image time lapse sequences of those axonemes which are clean, clamped at one end with proper beating, having $ 80$ to $90 \% $ of the filament in focus at a $ \Delta Z $ of $\rm 1.4~ \upmu m $ and do not have another reactivated filament in the surrounding area of $\rm 30~\upmu m ~\times~30~ \upmu m$.  The frame rate in the high speed imaging is suitably chosen to capture approximately 1000 beat cycles per axoneme, with $\rm \sim 55$ to $75$ conformations per beat cycle, for example 1200 frames per second for the axoneme in movie S1, whose beat frequency is approximately 16.63 Hz. Furthermore, the theoretically computed flow field is also depth averaged over this $ \Delta Z $ for appropriate comparison with the experimental flow field.

\vskip 2.5mm

\begin{figure}[h]
	\centering 
	\includegraphics[width=0.95\linewidth]{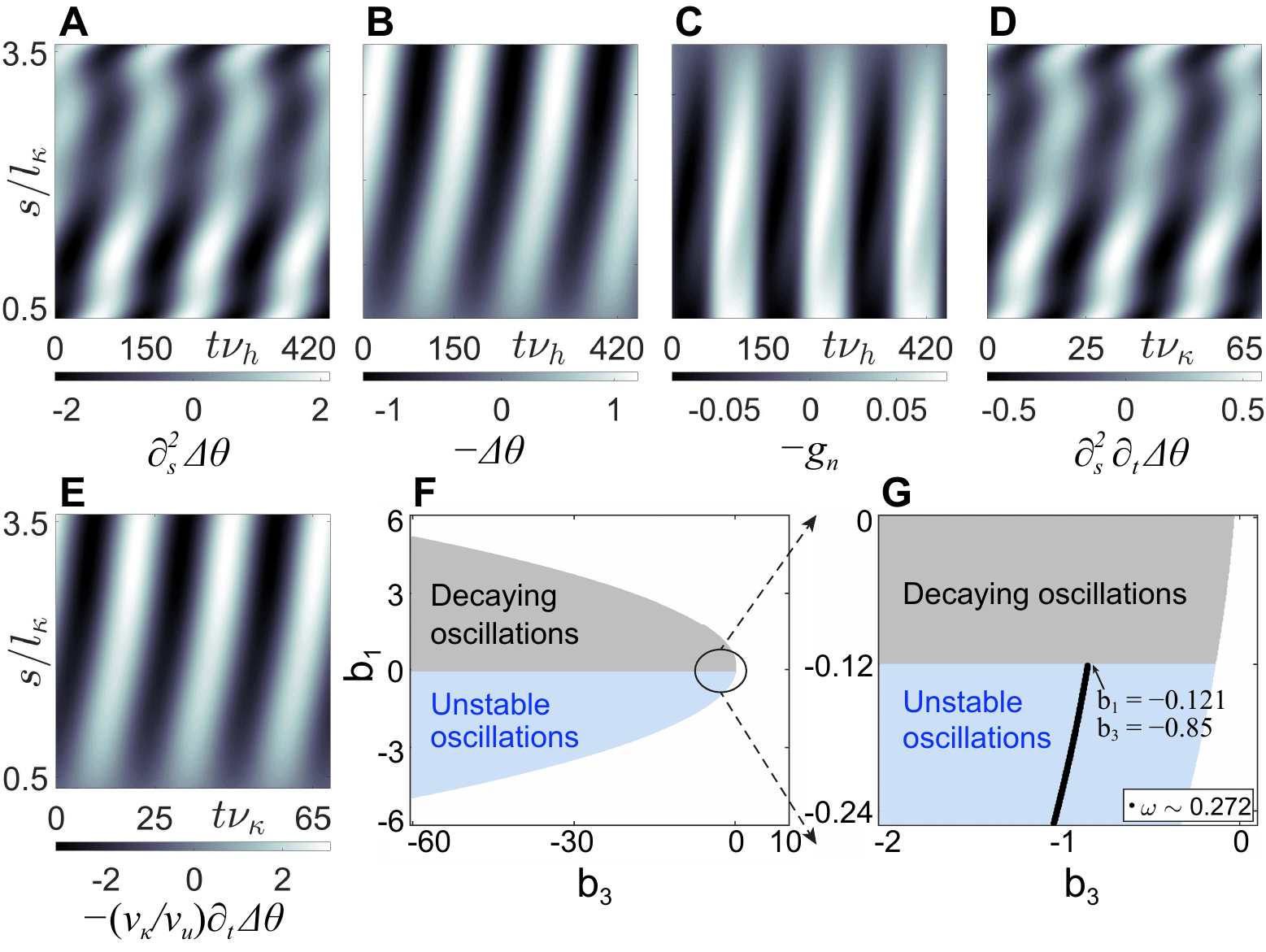} 
	\caption{\textbf{Dynamics and linear stability analysis.} Space-time plots of passive moments per unit length contributing to (\textbf{A}) bending elasticity, (\textbf{B}) shear elasticity and (\textbf{C}) normal component of the stress resultant, which includes external viscous drag; (\textbf{D}) bending friction and (\textbf{E}) shear friction. Scaling parameter, $EI/l_\kappa^{2}=80~{\rm pN}$. (\textbf{F}) Oscillatory nature of the complex frequency, $ z $, from linear stability analysis of Eq. \ref{coupled_eqn} in the parameter space ($ b_3,b_1 $) for the fundamental mode, $ q_1 $.  No oscillatory solutions exist in the white patches. (\textbf{G}) Zoomed-in view of (F) near the instability threshold, where $\omega$ denotes the oscillation frequency.}
	\label{Figure4-DynamicsStability} 
\end{figure}

\textbf{Extracting filament conformation from images:} 
We manually identify position coordinates of the axoneme for three beat cycles from the recorded time lapse sequences, which are then smoothened along $ s $ and $ t $ by using Savitzky-Golay filter of orders 3 and 5, respectively. The filament positions $ \boldsymbol{R}(s) $, arc length $ s $ are  nondimensionalized with the filament length $ L $ and time $ t $ with beat period $ T $, followed by setting the clamped end of the filament to (0,0). These nondimensionalized experimental positions are used to construct their Chebyshev interpolant as mentioned in the Results section. Higher order derivatives of the interpolation lose accuracy at the ends of the axoneme and data from those parts are, accordingly, discarded (fig.~S2). Hence, we consider $s/L \in [0.1, 0.9]$ when computing derivatives higher than first order and neglect the end values (as shown in Fig.~\ref{Figure4-DynamicsStability}, A to E).

\vskip 2.5mm

\textbf{Particle Tracking Velocimetry (PTV):} 
The recorded movies are background subtracted, before tracking the
tracer displacements, to consider only those tracers which are not
stuck to the coverslip and are following the flow. We track the tracer displacements in between two filament conformations with an infinitesimal time gap, $ \Delta t $ (for example, $ \Delta t \approx 3.33$ ms for power stroke waveforms and $ \Delta t \approx 4.98 $ ms in recovery stroke waveforms).  As the filament has a nonuniform beat frequency $\nu_b = 16.63 \pm 0.62$~Hz, over $\sim 940$ beat cycles, each conformation will not exactly repeat itself after one beat period. Hence, we cross-correlate each filament conformation for a given beat period with the whole recorded sequence of $\sim$ 940 beat cycles. The correlation peaks indicate the frames that have similar conformation. The stack of matched conformations is checked manually to delete any conformation with more than $10\%$ dissimilarity. The displacement of tracers in between these two conformations is calculated using standard MATLAB tracking routines and velocity vectors are obtained from $\sim500$ to $700$ beat cycles. The resulting velocity vectors are placed on $29\times29$ mesh grid, each of size $ \rm 0.74~\upmu m~\times~0.74~ \upmu m $ over the image, and
the mean at each grid point is computed. The gridded velocity vectors
are further smoothened using a $2\times2$ point averaging filter.

\vskip 5mm

\section*{Supplementary Materials}

It includes the following:\\
Section S1. Tension forces in the filament\\
Section S2. Linear stability analysis\\
Section S3. Why strain softening and shear thinning facilitates the instability to oscillations?\\
Section S4. Comparison with microscopic load dependent detachment model of dynein motors\\
Section S5. Estimates of $ \Gamma_u $ from literature\\
Movie S1. Movie of an isolated and reactivated axoneme in presence of tracers.\\
Table S1. Possible values of shear friction coefficient from the literature.\\
Fig. S1. Active filaments driven by slip at boundary vs driven internally by motors.\\
Fig. S2. Chebyshev differentiation of traveling wave parameters.\\
Fig. S3. Tension forces in the filament.\\
Fig. S4. Existence of oscillations with varying sliding friction coefficient.\\
References \cite{GangaRamaPRF,Saintillan2019Nonlinear,YokotaMabuchi1994,BaylyWilson2014interdoublet}.

\vskip 12mm

%%%%%%%%%%-----------------------------------------------
%%%%%%%%%%-------------Acknowledgements----------------
%%%%%%%%%%-----------------------------------------------

\noindent \textbf{Acknowledgements:} 
D.M. and P.S. thank J. Sutradhar, K. Ghosh, M. Jain, G. Prasath,
R. Govindarajan, T. G. Murthy, and D. Ghosh for useful discussions; B. J. Rao and S. Upadhyaya for help in growing \textit{Chlamydomonas}; K.-i. Wakabayashi for help in purifying
cilia; and S. Ramaswamy for important discussions and critical reading of our manuscript. D.M. and P.S. thank A. Baskaran for crucial inputs during the late stages of this work and
continuous support during peer review process. R.A. thanks R. E. Goldstein for helpful discussions and suggesting appropriate references. We thank A. Laskar for assistance in
the initial stages of this work and for sharing codes and data on phoretic particle chains.

\noindent \textbf{Funding:} 
This work was supported by the DBT/Wellcome Trust India Alliance Fellowship (grant number IA/I/16/1/502356) awarded to P.S. R.A. acknowledges support from the Isaac Newton Trust.

\noindent \textbf{Author Contributions:} 
R.A. and P.S. designed the research, D.M.
performed the research, and all authors wrote the paper.

\noindent \textbf{Competing Interests:} 
The authors declare that they have no competing interests.

\noindent \textbf{Data and materials availability:} All data needed to evaluate the conclusions in the paper are present in the paper and/or the Supplementary Materials. Additional data related to this paper may be requested from the authors.

%%%%%%%%%%------------ BIBLIOGRAPHY --------------------
%%%%%%%%%%-----------------------------------------------

%%%% To change the longbibliography class to include only initials of authors
\makeatletter
\def\@bibdataout@aps{%
	\immediate\write\@bibdataout{%
		@CONTROL{%
			apsrev41Control%
			\longbibliography@sw{%
				,author="08",editor="1",pages="1",title="0",year="1"%
			}{%
				,author="08",editor="1",pages="1",title="",year="1"%
			}%
		}%
	}%
	\if@filesw \immediate \write \@auxout {\string \citation {apsrev41Control}}\fi 
}
\makeatother

\bibliography{ciliaref}

%%%---------------FIGURES--------------------
%%%--------------------------------------------------
%%%%%%-------------------------------------------------
\clearpage \newpage

%%%%%%%--------------------------------------------------------
%%%%%%-----------------------------------------------------
%%%%%%%%%%%%%%%%%%%%%%%%%%%%%%%%%%%%%%%%%%%%%%%%%%%%%%%%%%%%%%
%%%%%------------------------------------------------------------
%%%%%%-------------Supplementary information--------------------

\clearpage
\newpage

\begin{center}
\section*{\large Supplementary Materials}
\end{center}

%%%%--------------For supplementary----------------
\newcommand{\beginsupplement}{
	\setcounter{equation}{0}
	\renewcommand{\theequation}{S\arabic{equation}}
	\setcounter{figure}{0}
	\renewcommand{\thefigure}{S\arabic{figure}}
	\setcounter{table}{0}
	\renewcommand{\thetable}{S\arabic{table}}
}
%%%%%%%%---------------------------------------------

\beginsupplement

\subsection*{{\normalsize Section S1. Tension forces in the filament}}

The tangential component of the stress resultant is the tension within the filament \cite{GangaRamaPRF}, which we can compute from filament velocity as $ F_t(s)=\mathbf{t}(s) \cdot \int_s^L \mathbf{f}^v(s') ds' = -\mathbf{t}(s) \cdot \int_s^L \mathbf{\gamma} \cdot \dot{\mathbf{R}} (s')ds' = -\gamma_n g_t(s) $, where, $ g_t(s) = \mathbf{t}(s) \cdot \int_s^L [\dot{R}_n(s')\mathbf{n}(s') + (\dot{R}_t(s')/2)\mathbf{t}(s') ] ds' $. The tension force in the filament is also too small (fig.~S3) compared to the internal elastic forces.

\subsection*{{\normalsize Section S2. Linear stability analysis}}

We Fourier transform the coupled equations of motion (Eq. \ref{coupled_eqn} in main text) in space and time with the following convention
\begin{equation} \nonumber
\Delta\uptheta(s,t) = \int \frac{dz}{2\pi} \sum_n \widetilde{\Delta\uptheta}(q_n,z)e^{i(q_ns-z t)},
\quad \text{[similarly for $ m^\mathcal{A}(s,t) \to \widetilde{m^\mathcal{A}}(q_n,z) $]}
\end{equation}
where $ q_n $ is the discretized wavenumber of the $n$th mode due to finite length of the filament and $ z $ is the complex frequency. In matrix form,
\begin{equation}\label{1b_matrix}
\begin{bmatrix}
(1+q_n^2) -iz (q_n^2+ \upnu_\upkappa/\upnu_u) & -1  \\
b_3 & iz-b_1
\end{bmatrix}
\begin{bmatrix}
\widetilde{\Delta\uptheta} \\ \widetilde{m^\mathcal{A}}
\end{bmatrix}
=
\begin{bmatrix}
0 \\ 0
\end{bmatrix}
\end{equation}
For non-trivial solution, the determinant of the above matrix must be zero. Hence the dispersion relation is
\begin{equation} \label{dispersion_reln}
\big[(1+q_n^2)-iz (q_n^2+ \upnu_\upkappa/\upnu_u)\big](iz-b_1) + b_3 = 0
\end{equation}
This is a quadratic equation in the complex frequency $ z $ of the form $ A(q_n)z^2 + B(q_n)z +C(q_n)=0 $ whose coefficients are
$A(q_n) = q_n^2 + \upnu_\upkappa/\upnu_u$,
$B(q_n) = i \big[(1+q_n^2) + b_1(q_n^2 +\upnu_\upkappa/\upnu_u) \big]$ and
$C(q_n) = b_3 - b_1(1+q_n^2)$.
Roots are $ z_{1,2}=\big[-B(q_n) \pm \sqrt{B^2(q_n) - 4A(q_n)C(q_n)}\big]/2A(q_n) $.
As the time dependent component in the solution is $ e^{-izt} $, existence of the real part of the root will imply that the solution is oscillatory and sign of the imaginary part of $ z $ will decide if the solution is growing or decaying. Hence, the conditions for unstable oscillations are $ Im[z] >0 $ and $ Re[z] \neq 0 $. The frequency of the oscillations is therefore given by $ \upomega =-Re[z] $.

\subsection*{{\normalsize Section S3. Why strain softening and shear thinning facilitates the instability to oscillations?}}

We Fourier transform the dynamical equation of motion (Eq. \ref{torque_eqn2_ND} in the main text) in space and time under oscillating shear of fundamental frequency $ \upomega $ i.e. $ \Delta\uptheta \sim \widetilde{\Delta\uptheta}e^{i(q_n s+\upomega t)}  $ where $ \upomega = -Re(z) $ is real and $ q_n $ is the discretized wavenumber of the $ n $th mode.
\begin{equation}
	( -q_n^2 -i\upomega q_n^2 - 1 - i \upomega\upnu_\upkappa/\upnu_u) \widetilde{\Delta\uptheta} + \widetilde{m^\mathcal{A}}=0
\end{equation}
Replacing the fundamental Fourier mode of active stress as $ \widetilde{m^\mathcal{A}}= (G'+i\upomega G'')\widetilde{\Delta\uptheta}$, where where $ G' $ and $ G'' $ corresponds to elastic and viscous response of the system,  respectively and are related to $ b_1,b_3 $ such that $ b_1,b_3<0 \implies G', G''>0 $ (see main text),  we obtain
\begin{equation}
	\underbrace{-(1+q_n^2)\widetilde{\Delta\uptheta}}_{\text{passive elastic}} \underbrace{-i\upomega(q_n^2 +\upnu_\upkappa/\upnu_u)\widetilde{\Delta\uptheta}}_{\text{passive viscous}} + \underbrace{(G'+i\upomega G'')\widetilde{\Delta\uptheta}}_{\text{active elastic + viscous}}=0
\end{equation}
Hence, the passive elastic and viscous terms in the equation of motion are negative i.e. they resist the sliding caused by the active dynein motors. Now, if $ G',G''<0 $, the system's passive spring constant and friction coefficient get renormalized by the ATP dependent dynein activity and thus, both the active and passive components of the system resist the sliding ultimately leading to a quiescent stable state of the filament. On the other hand, if $ G',G'' >0 $, dynein motors work against the material response so that the system becomes unstable and undergoes oscillations. Here, $ G' > 0 $ indicates that elastic stresses reduce within the axoneme as motor activity increases which is called `strain softening' and $ G''>0 $ indicates that the axoneme becomes less viscous  with increasing motor activity which is called `shear thinning'.

We note that nonlinear viscoelastic effects are not needed for an active material, such as the axoneme, to  shear thin/strain soft, because the nonequilibrium active stresses generated by the dynein motors can produce structural rearrangements within the axoneme by binding and unbinding the dynein cross-bridges across microtubule doublets. This is in contrast to passive equilibrium systems which have to be intrinsically nonlinear to strain soft/shear thin under large external shear as thermal energy alone is insufficient to drive structural rearrangements in such systems.

\subsection*{{\normalsize Section S4. Comparison with microscopic load dependent detachment model of dynein motors}}

In the microscopic load dependent detachment model, the motor detachment rate ($ k_{off} $) is assumed to increase exponentially with increasing load (i.e. single motor force, $ f_+ $), which in turn decreases linearly with the dynein sliding speed ($ v_d $) by the force-velocity relationship of the motors \cite{Riedel-Kruse2007,Oriola2017Nonlinear,Saintillan2019Nonlinear}.
\begin{equation}
	k_{off} (f_+) = k_0 \exp\bigg[\frac{f_+}{f_c} \bigg] = k_0 \exp \bigg[\frac{\bar{f}-f'v_d}{f_c} \bigg]
\end{equation}
where $ \bar{f} $ is the dynein stall force, $ f' $ is the slope of the dynein force-velocity curve, $ f_c $ is the characteristic unbinding force generally given by $ f_c \approx \bar{f}/2 $ \cite{Riedel-Kruse2007,Oriola2017Nonlinear}. The force-velocity slope is given by $ f'=\bar{f}/v_0 $, where $ v_0 $ is the dynein velocity at zero load. In the limit of low sliding speed i.e. $ \dfrac{f' v_d}{f_c} << 1 $, the above exponential relation linearizes to
\begin{equation}
	k_{off} (f_+) = \bar{k}_{off} \bigg[ 1- \frac{f' v_d}{f_c}\bigg]
\end{equation}
where $ \bar{k}_{off} = k_{off}(\bar{f}) = k_0 \exp(\bar{f}/f_c) $ is the motor detachment rate at stall (refer to Eq. B4 in Appendix B of \cite{Riedel-Kruse2007}). Now let us see if our experiments near the critical ATP concentration satisfy the linearizing condition, $ \dfrac{f' v_d}{f_c} << 1 \implies \dfrac{v_d}{v_0/2} << 1 $.\\
Axoneme being a cross-linked filament, the angular speed of the axoneme ($ \partial_t\Delta\uptheta $) is related to the sliding speed per dynein motor as $ v_d = a_{MT} \partial_t\Delta\uptheta/L \uprho \bar{p} $ \cite{Riedel-Kruse2007}). Here $ a_{MT} =24$ nm is the MT interdoublet spacing in which the dyneins work \cite{Nicastro2006}, $ L $ is the filament length at which angular speed is calculated from experiments, $ \uprho = 198~\upmu$m$ ^{-1} $ is motor density \cite{Nicastro2006} and $ \bar{p} =0.02 $ is the fraction of motor domains that are attached to MT, equivalent to the duty ratio \cite{Howard2001mechanics,Oriola2017Nonlinear}. Therefore, $ L\uprho \bar{p} $ is the total number of motors bound to a single MT along the length of the filament. At 60 $ \upmu $M ATP, $ \partial_t \Delta\uptheta \approx 80 $ rad/s (from Fig.~\ref{Figure4-DynamicsStability}E of main text) at $ L \approx 9~\upmu $m. Using these values $ v_d \approx 54 $ nm/s. Earlier experiments have measured the zero load dynein velocity at 60 $ \upmu $M ATP to be $ v_0 \approx 2 ~\upmu$m/s \cite{Howard2001mechanics,YokotaMabuchi1994}. Hence the ratio of dynein sliding speed in our experiments to the half of its zero load velocity $ \dfrac{v_d}{v_0/2} \approx 0.05 << 1 $. To summarize, the axoneme beating near the instability threshold at 60 $ \upmu $M ATP is in the dominantly linear regime (equivalently weakly nonlinear regime) of the post-bifurcation dynamics. Therefore, our choice of linear constitutive relationship for the active moment and the associated linear stability analysis is valid near the instability threshold (also refer to Fig.~3a and associated text of \cite{Oriola2017Nonlinear}).

Now, that we have shown our experiments are consistent with the condition for linearizing the exponential dependence of motor detachment rate on sliding speed, we connect our constitutive equation for active moment to this linearized version of microscopic motor dynamics model (\textit{sliding control motor coordination model}) proposed by Riedel-Kruse et al. \cite{Riedel-Kruse2007}.
In  \cite{Riedel-Kruse2007}, the active shear force per unit length is related to the shear displacement by a response function, $ \upchi=K+i\upomega \uplambda $, as per their notation. The sign convention of $ G' $ and $ G'' $ is opposite to the elasto-viscous response coefficients ($K$ and $\uplambda$) of  \cite{Riedel-Kruse2007} (also of \cite{CamaletJulicher2000,CamaletJulicherProst1999}), as active drive in these references has opposite sign convention. The equivalence of our response coefficients to their microscopic  model of load dependent detachment of motors (refer to Eq. B9 in Appendix B of  \cite{Riedel-Kruse2007}) are as follows:\\
(a) $ G'\equiv -a^2 K = 2a^2 \uprho \bar{f} \frac{f'}{f_c}\bar{p}(1-\bar{p}) \frac{\upomega^2 \bar{\uptau}}{1+(\upomega \bar{\uptau})^2}$. This quantity is always positive. \\
(b) $ G'' \equiv -a^2\uplambda = -2a^2 \uprho f'\bar{p} \big[ 1 - \frac{\bar{f}}{f_c} \frac{(1-\bar{p}) }{\{1+(\upomega \bar{\uptau})^2\}} \big] $. This quantity can be positive or negative.\\
In the above expressions, $ \bar{\uptau} $ is the relaxation time of motor attachment/detachment and all other variables are already defined in the preceding paragraph. The sign in (b) depends on $ \bar{\uptau} $ and $ \bar{p} $ as:  (i) $ G''>0 $ for $ \upomega \bar{\uptau}<<1 $ and $ \bar{p} \sim 0 $ and
(ii) $ G''<0 $ for $ \upomega \bar{\uptau}>>1 $ and $ \bar{p} \sim 1 $.
For our active filament $ G''>0 $ i.e. $ b_3<0 $ asserts that $ \upomega \bar{\uptau}<<1 $ and $ \bar{p} \sim 0 $. This means that the axonemal dynein motors are short lived with low duty ratio, which agrees with \cite{Howard2001mechanics}.

\subsection*{{\normalsize Section S5. Estimates of $ \Gamma_u $ from literature}}

All parameters, except the shear friction coefficient, have been experimentally measured for an axoneme or at least for microtubules in the existing literature as mentioned in the main text. There is uncertainty in the value of $ \Gamma_u $. Following (table \ref{Table:SlidingFriction}) are the values used for this coefficient for constructing active filament models in the literature, except for the last entry which is an experimental study on shear elasticity.

We note that the exact value of this coefficient does not affect the existence of oscillation in our theoretical model (fig.~S4) rather it modulates the magnitudes of the viscoelastic response coefficients of the active stress. The sliding friction coefficient $ \Gamma_u $ appears in the dimensionless dynamical equation as the ratio $ \upnu_\upkappa/\upnu_u $. Variation of $ \upnu_\upkappa/\upnu_u $ implies variation in $ \Gamma_u $ because
all parameters except $ \Gamma_u $ are experimentally known for an axoneme/MT. Figure~S4 shows that unstable oscillations exists for $ \upnu_\upkappa/\upnu_u \in [0.1,~ 50]$.  Correspondingly, the varation of $ \Gamma_u $ in this range is $[10^{-7},~ 0.5 \times 10^{-4}] $ Ns/m.

\begin{table}[h]
	\begin{center}
		\begin{tabular}{|c|c|c|}
			\hline
			Reference & $ \Gamma_u $ [$ \times 10^{-6} $] & Comparison of shear friction with elasticity \\
			& Ns/m &\\
			\hline \hline
			Brokaw \cite{Brokaw1972computer} & 0.06-0.51 & negligible compared to elastic terms\\
			\hline
			Murase-Shimizu \cite{murase1986model}  & 30  & very high compared to elastic terms,\\
			& & overdamping the system\\
			\hline
			Bayly-Wilson \cite{BaylyWilson2014interdoublet} & 0.5 & negligible compared to elastic terms\\
			\hline
			Bayly-Dutcher \cite{BaylyDutcher2016Flutter} & 0.16 & negligible compared to elastic terms\\
			\hline
			\textbf{Ingmar Riedel} \cite{riedel2005mechanics}  & \textbf{10} & \textbf{comparable and competing with elastic terms, \checkmark}\\
			\hline
			Minoura-Yagi- & 365 & very high compared to elastic terms,\\
			Kamiya \cite{Minoura2001Spring} & &   overdamping the system{*}  \\
			\hline
		\end{tabular}
		\\{*}Obtained from digitizing Fig. 5 of \cite{Minoura2001Spring}, and fitting a creep response function for Kelvin-Voigt model of viscoelasticity.
		\caption{\textbf{Possible values of shear friction coefficient from the literature.} Shear friction coefficient from existing literature and reasons for neglecting/accepting them.}
		\label{Table:SlidingFriction}
	\end{center}
\end{table}

\section*{Movie caption}

\textbf{Movie S1.} \textbf{Movie of an isolated and reactivated axoneme in presence of tracers.} High speed phase contrast movie of a reactivated and clamped \textit{Chlamydomonas} axoneme at 60 $\upmu$M ATP in presence of 200 nm tracer particles. Scale bar, 5 $\upmu$m.

%%%%%%----------------SM Figures-------------------------------
%%%%%%---------------------------------------------------------

\clearpage

\begin{figure}[t]
	\centering
	\includegraphics[width=0.9\linewidth]{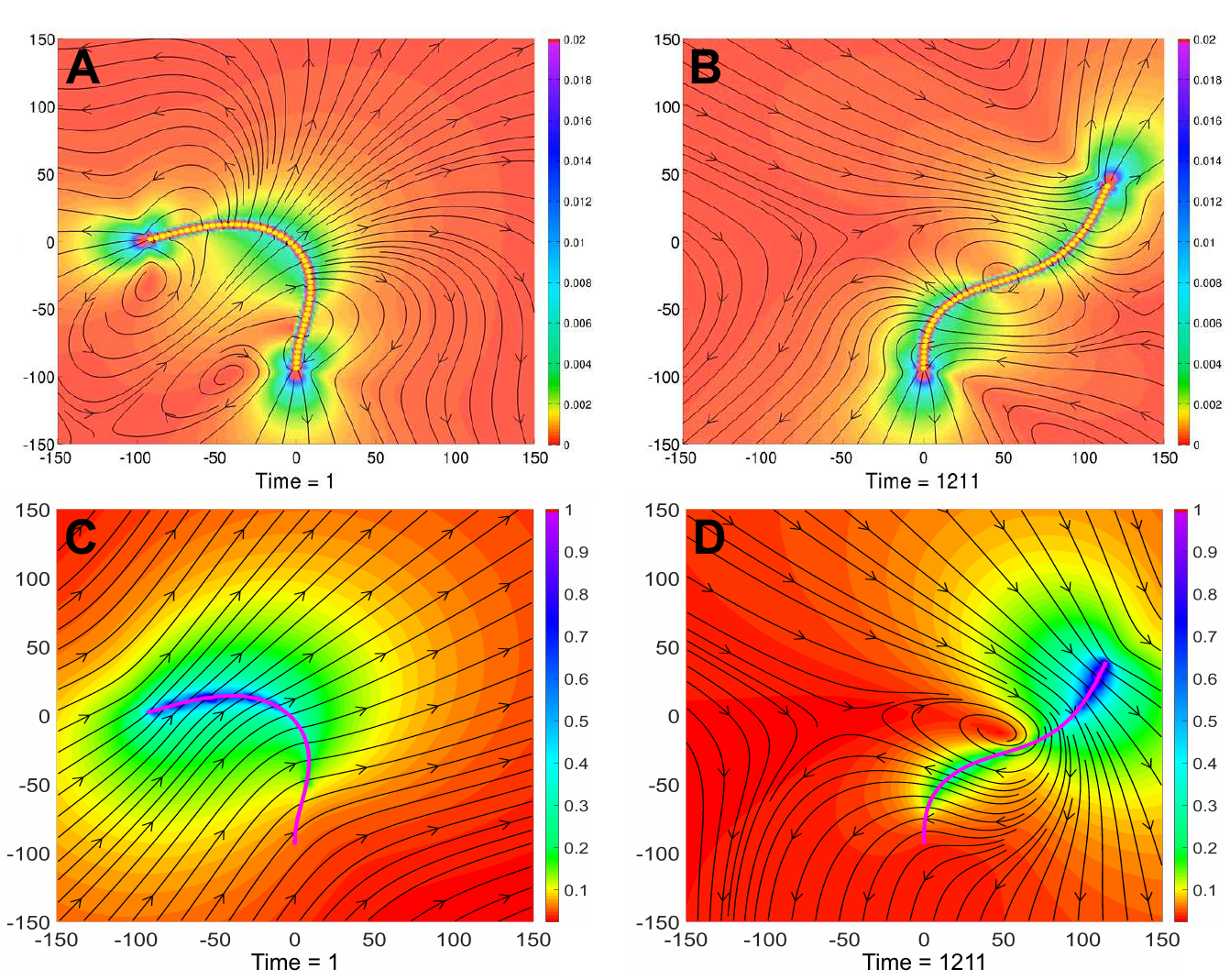}
	\caption{\textbf{Active filaments driven by slip at boundary vs driven internally by motors.} (\textbf{A} and \textbf{B}) Flow fields of a planar flexible beating of a clamped filament, consisting of chemomechanically active beads, at two instants of the oscillation cycle, adapted from Supplementary video 2 in \cite{LaskarAdhikari2013SciRep}. (\textbf{C} and \textbf{D}) Computed flow fields using slender body approximation and resistive force theory (unbounded flow) where the filament positions were extracted from the video. The mismatch between (A and C) and (B and D) imply that the filament must not be internally driven instead slip driven as expected of a phoretic chain. The colorbars to the right of (C) and (D) represents the normalized speed.}
	\label{FigS1_ActiveVsDriven}
\end{figure}

\begin{figure}[t]
	\centering
	\includegraphics[width=0.85\linewidth]{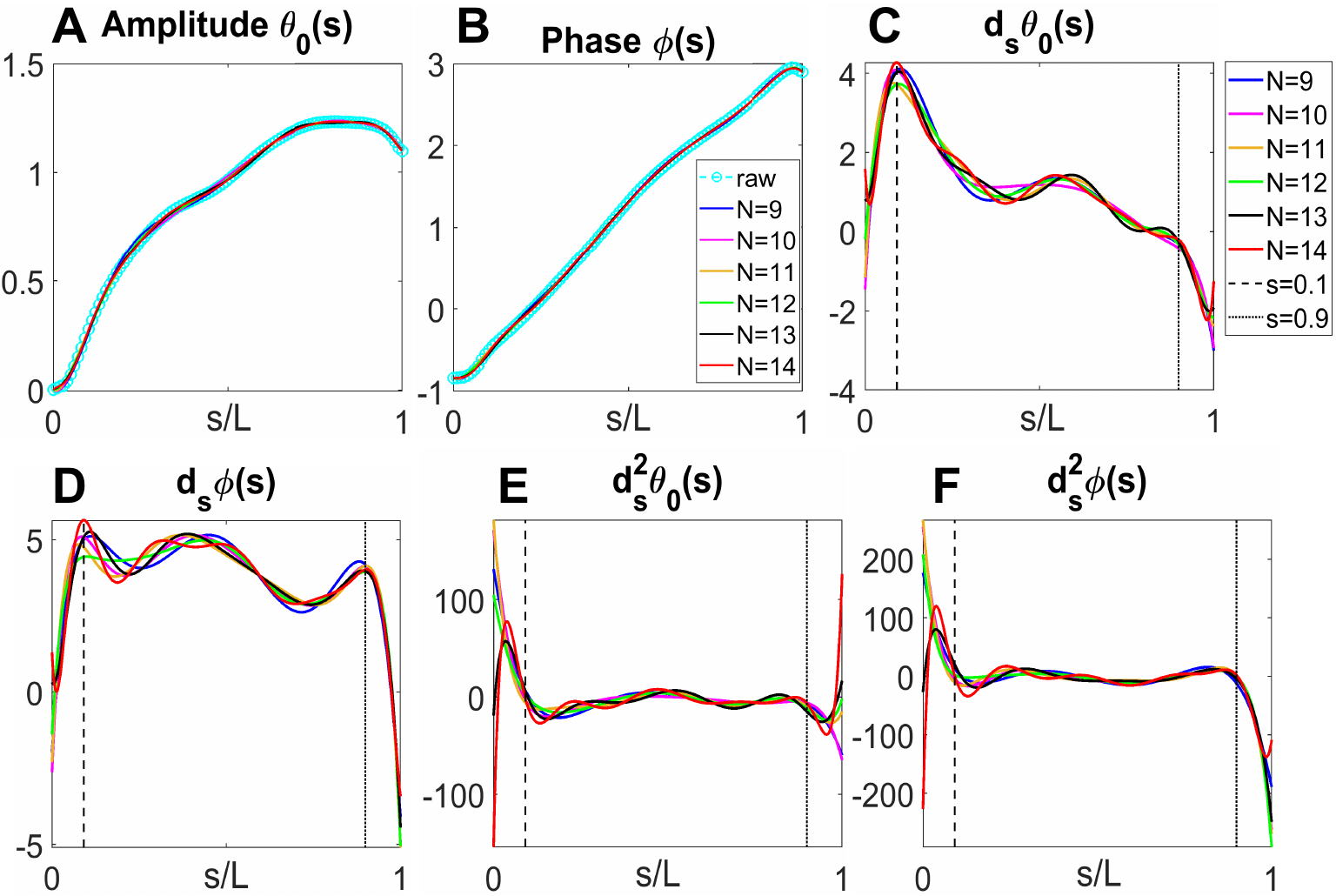}
	\caption{\textbf{Chebyshev differentiation of traveling wave parameters.} (\textbf{A}) Amplitude ($\uptheta_0$) and (\textbf{B}) phase ($\upphi$) of the traveling wave parameterization to $ \uptheta $ are plotted in cyan circles. Interpolation to them for different Chebyshev polynomial orders $ N $ almost converges to the actual value.  First order Chebyshev differentiation of (\textbf{C}) $\uptheta_0$ and (\textbf{D}) $\upphi$ with respect to $ s $ for different polynomial orders. Second order Chebyshev differentiation of (\textbf{E}) $\uptheta_0$ and (\textbf{F}) $\upphi$ with respect to $ s $ for different polynomial orders. Legends of (C to F) are shown beside (C).}
	\label{FigS2_ChebyshevDer}
\end{figure}

\begin{figure}[t]
	\centering
	\includegraphics[width=0.45\linewidth]{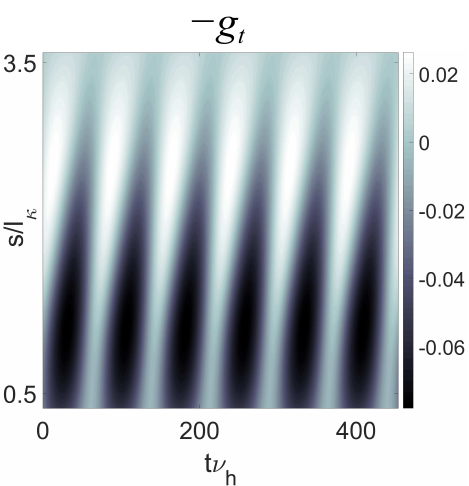}
	\caption{\textbf{Tension forces in the filament.} Space-time plot of the tension force i.e. tangential component of the stress resultant, along the filament for three beat cycles. The length and time scales are $ l_\upkappa $ and $ 1/\upnu_h $. The colorbar represents its magnitude, with the force scale $ EI/l_\upkappa^{2}=80~{\rm pN} $.}
	\label{FigS3_Tension}
\end{figure}

\begin{figure}[h]
	\centering
	\includegraphics[width=0.58\linewidth]{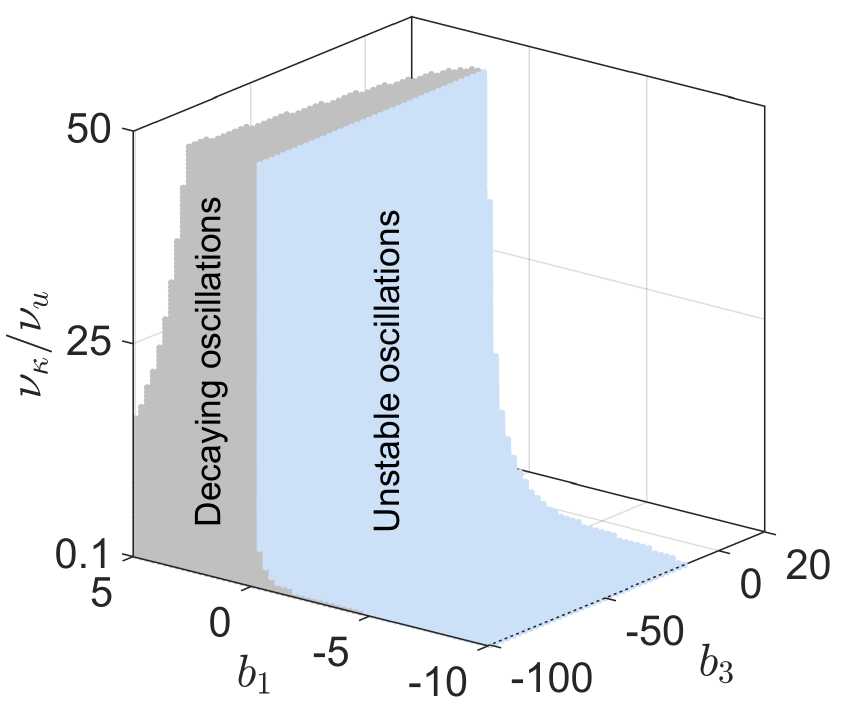}
	\caption{\textbf{Existence of oscillations with varying sliding friction coefficient.} Oscillatory nature of the complex frequency, $ z $ in the parameter space ($ b_3,b_1,\upnu_\upkappa/\upnu_u $) for the fundamental mode, $ q_1 $.  No oscillatory solutions exist in the white regions. The range of $ \upnu_\upkappa/\upnu_u \in [0.1,~ 50] $ corresponds to $ \Gamma_u \in [10^{-7},~ 0.5 \times 10^{-4}] $ Ns/m (considering the value of $ \upnu_\upkappa $ to be fixed at 375 Hz). The blue region of unstable oscillations for all these values of $ \Gamma_u $ indicate that the magnitude of $ \Gamma_u $ does not affect the existence of oscillations.}
	\label{FigS4_GammaU}
\end{figure}

\end{document}